\documentclass[12pt]{article}
\usepackage[a4paper,margin=2cm]{geometry}
\usepackage[T1]{fontenc}
\usepackage{times}
\usepackage{enumerate}
\usepackage{amsmath,amssymb,amsthm,amsfonts}
\usepackage{graphicx}
\graphicspath{ {./images/} }
\usepackage{authblk}
\usepackage{color}
\usepackage[all,cmtip]{xy}
\usepackage{cite}
\usepackage{bm}
\usepackage{hyperref}
\usepackage{marginnote}
\usepackage{booktabs}
\usepackage{threeparttable}
\usepackage{bbm}
\usepackage{caption}
\usepackage{rotating}
\usepackage{float}
\newcommand{\Esp}[2][]{\mathbb{E}_{#1}\!\left[ #2 \right]}

\newcommand{\Prob}[2][]{\mathbb{P}_{#1}\!\left( #2 \right)}

\newcommand{\indi}[1]{\mathbf{1}\left( #1 \right)}
\DeclareMathOperator*{\Cov}{Cov}
 
\title{Evaluating the influence of treatment-effect heterogeneity on discrimination}

\author[1]{Florie Bouvier %\thanks{Correspondence to: Florie Bouvier (\href{mailto:florie.brion-bouvier@u-paris.fr}{florie.bouvier@u-paris.fr})\\
}
\author[1]{Etienne Peyrot}
\author[1]{Fran\c{c}ois Petit}
\author[1,2]{Rapha\"el Porcher}

\affil[1]{Universit\'e Paris Cit\'e and Universit\'e Sorbonne Paris Nord, Inserm, INRAE, Center for Research in Epidemiology and StatisticS (CRESS), F-75004 Paris, France
}
\affil[2]{Centre d'\'Epid\'emiologie Clinique, Assistance Publique-H\^opitaux de Paris, H\^otel-Dieu, Paris, France}

\date{}
\begin{document}
	
\theoremstyle{plain} % style plain
\newtheorem*{theorem*}{Theorem} % unnumbered theorem
\newtheorem{theorem}{Theorem}[section]
\newtheorem*{corollary*}{Corollary} % unnumbered corollary
\newtheorem{corollary}[theorem]{Corollary}
\newtheorem{proposition}[theorem]{Proposition}
\newtheorem*{lemma*}{Lemma} % unnumbered lemma
\newtheorem{lemma}[theorem]{Lemma}
\theoremstyle{definition} % style definition
\newtheorem{definition}[theorem]{Definition}
\newtheorem{example}[theorem]{Example}
\newtheorem{remark}[theorem]{Remark}
\newtheorem{assumption}{Assumption}
\newtheorem{examples}[theorem]{Examples}
\newtheorem{question}[theorem]{Question}
\newtheorem{Rem}[theorem]{Remark}
\newtheorem{Notation}[theorem]{Notations}

\counterwithout{equation}{section}

\maketitle
\begin{abstract}
Analyzing the heterogeneity of treatment effects is crucial in personalized medicine to identify which patients will benefit from specific treatments. The performance of a conditional average treatment effects model to guide treatment decisions can be assessed in different ways, with an important one being the model's ability to effectively discriminate between individuals who benefit from the treatment and those who do not. While many methods and algorithms have been proposed to develop conditional average treatment effects models and individualized treatment rules, little is known about the discriminative ability that can be achieved according to the population's underlying distribution of treatment effects. In this work, we computed the  discrimination  that can be achieved under oracle CATE for a panel of 20 distributions with varying average treatment effects and levels of heterogeneity. The assessment included the following discrimination metrics: the c-statistic for benefit, the concentration of benefit, and the population average prescriptive effect (PAPE). Results showed that the three metrics employed in this study did not require the same levels of treatment effect heterogeneity to lead to high discrimination results. Notably, achieving high c-statistic for benefit and PAPE values required greater heterogeneity than obtaining high concentration of benefit values. The three metrics considered behave very differently across the distributions. For instance, the concentration of benefit can indicate perfect discrimination in settings with negligible treatment-effects heterogeneity.

\end{abstract}

\section{Introduction}\label{sec:intro}

The analysis of heterogeneity of treatment effects, which represents the non-random variability in the direction or magnitude of a treatment effect, is pivotal in personalized medicine for identifying patients who would benefit from specific treatments. This step facilitates the development of individualized treatment rules (ITRs), which are decision rules recommending treatment based on individual patient characteristics \cite{VARADHAN2013818, kent_personalized_2018, kent_predictive_2020}.
\\
Extensive literature is available for evaluating the performance of risk prediction models \cite{book_harrell,book_steyerberg}, as various metrics are used to assess a model's performance comprehensively, focusing on aspects such as calibration \cite{hosmer1980goodness,miller1993validation} and discrimination \cite{harrell1982evaluating}. Further, the clinical utility of a model can be quantified using methods like net benefit and decision curve analysis \cite{vickers2006decision}. Demonstrating strong discrimination is considered a key aspect of a model's performance \cite{diamond_what_1992,gail_criteria_2005}.
\\
As opposed to risk prediction models, limited works exist on the performance of conditional average treatment effects (CATEs) models and ITRs. The assessment of ITR performance has been described by some metrics, including the concept of the ITR's value \cite{tsiatis_dynamic_2020}, the benefit of the rule in terms of assigned treatment \cite{Janes2014}, and other related metrics \cite{grolleau2023comprehensive}. These metrics primarily focus on evaluating the benefits of implementing ITRs. Conversely, fewer studies have concentrated on the performance of CATE models due to the inherent challenge in measuring such a performance, as the outcomes of both treatments are typically not observable within a single patient. Consequently, conventional risk prediction metrics cannot effectively quantify the performance of models predicting CATE. Calibration can be assessed by comparing predicted treatment benefits to estimated treatment benefits across deciles of CATE \cite{van_klaveren_models_2019}. Recent metrics have also been proposed to evaluate discrimination, which gauges the ability to differentiate individuals who benefit from those who do not. In this work, we specifically focused on three of them: the c-statistic for benefit \cite{van_klaveren_proposed_2018}, the concentration of benefit \cite{Sadatsafavi2020} and the population average prescriptive effect (PAPE) \cite{imai_experimental_2021}.
\\
Previous works have indicated that a significant disparity in risk of events is necessary to effectively differentiate between patients at high risk of developing the outcome and those at low risk \cite{pepe_limitations_2004,cook_use_2007}. The extent of heterogeneity in treatment effects required to derive effective ITRs with optimal discriminative ability has yet to be thoroughly investigated. Specifically, there is an interest in understanding the discrimination that an ITR can achieve under different distributions of conditional average treatment effects. 
\\
Our objectives were to assess the discriminative capacity of the three aforementioned metrics under diverse treatment effect distributions and to compare their relative discrimination performance.  We calculated, via numerical integration, the three discrimination metrics mentioned above for the CATE under different distributions of the treatment effects.
The remaining sections of the paper are structured as follows: Section \ref{sec:methods} provides details on the metrics used and how the distributions were defined, Section \ref{sec:res} showcases the results, and Section \ref{sec:discuss} concludes the paper with a discussion.

\section{Methods}\label{sec:methods}
This section introduces the metrics employed for the analysis and describes how they were numerically calculated. Twenty distributions featuring heterogeneous treatment effects, with alterations to the direction or magnitude of these effects, were chosen to illustrate the achievable discrimination. Three discrimination metrics were computed for all distributions (see subsection~\ref{ssec:metrics}). 
\\
\\

In Rubin's counterfactual framework, the individual benefit for individual $i$ is defined as $B_i = Y_i^1 - Y_i^0$ where $Y_i^{0}$ and $Y_i^{1}$ refer to the outcomes that would be observed if the individual $i$ was assigned to either the control or the experimental treatment \cite{rubin_estimating_1974}. 
In practice, $B_i$ is not observable. Hence, we resort to covariates $X$ to estimate the so-called conditional average treatment effects $\tau(X) = \mathbb{E}(Y^{1} - Y^{0}|X)$ which represents the expected difference in outcomes of two treatments for individuals with given characteristics.
In this project, we focus on binary outcomes. Without loss of generality, we assume that $Y=1$ is a desirable event so that $\tau > 0$ would lead to recommend the experimental treatment.

\noindent In this project, we position ourselves in an oracle situation where the CATE is known, to remove variations induced by the estimation step required in practice.

\subsection{Metrics}\label{ssec:metrics}

\noindent The discriminatory performance of the CATE i.e. the ability of an ITR to separate individuals who benefit from taking the treatment from those who do not,  was measured by the c-statistic for benefit, the concentration of benefit, and the population average prescriptive effect.\\
\\
The original c-for-benefit is a matching-based estimator of a pairwise concordance between predicted treatment benefit and observed treatment benefit. Because individual causal benefit is not jointly observable in trial data, the original implementation uses matched treated-control pairs and defines observed benefit as the within-pair outcome difference. In our oracle setting, both potential outcome risks are specified by the data-generating mechanism, so no matching approximation is required. We therefore consider the corresponding individual-level oracle estimand. This leads to the following definition.

\paragraph{c-statistic for benefit}
The c-statistic for benefit ($C_{\text{fb}}$) is the probability that, for two patients u and v with an unequal individual benefit, the patient with the higher individual benefit also has a higher predicted benefit \cite{van_klaveren_proposed_2018}. Higher values of the c-statistic for benefit indicate a higher performance. It can be mathematically expressed as:
$$C_{\text{fb}} = P\Big(\tau_u>\tau_v \,|\, B_u>B_v\Big)$$ 
where $B_u$ and $B_v$ represent the individual benefits of patients $u$ and $v$ and where $\tau_u$ and $\tau_v$ represent the predicted benefit of patients $u$ and $v$ respectively.\\

Some concerns have been raised regarding the quality of the $C_{\text{fb}}$ as a discrimination metric, notably recent studies have shown that the $C_{\text{fb}}$ is not a proper scoring rule or that it is sensitive to the way the pairs are matched \cite{hoogland_evaluating_2022,xia_methodological_2023}.

\paragraph{Concentration of benefit}
The concentration of benefit ($C_b$) metric evaluates the extent to which covariates effectively capture the variation in treatment effects and measures how well the CATE model identifies the individuals who experience greater benefits from the treatment \cite{Sadatsafavi2020}. Larger concentration of benefit values, in particular approaching $1$, are usually interpreted as indicating better discrimination. The concentration of benefit can be mathematically denoted as 

\begin{equation*}
C_b =
\begin{cases}
    1 - \frac{\Esp{\tau}}{\Esp{\max(\tau,\tau')}},\text{ if }\Esp{\tau} \geq 0,\\
    1 - \frac{\Esp{\tau}}{\Esp{\min(\tau,\tau')}}, \text{ otherwise.}
\end{cases}
\end{equation*}
where $\tau'$ is an independent copy of $\tau$.

\paragraph{Population average prescriptive effect}
The population average prescriptive effect (PAPE) is the difference in mean outcomes when comparing an ITR and a treatment rule that randomly administers treatment to the same proportion of patients \cite{imai_experimental_2021}. The PAPE evaluates the efficacy of the ITR in recommending treatment only to the individuals who benefit from it. It is expressed as $$\text{PAPE}=\Esp{Y(r)-p_r Y^1-(1-p_r)Y^0}$$ where $Y(r) = rY^{1} + [1-r]Y^{0}$ represents the outcome observed if treatment allocation follows the rule $r$ and $p_r$ denotes the proportion of patients assigned to the experimental treatment under the ITR $r$. It can be re-expressed as
\[
  \text{PAPE} = \mathbb{E}[(r-p_r)\tau].  
\]
The PAPE ranges from $-0.5$ to $0.5$, with higher PAPE values implying a better performance of the ITR. A value of $0$ indicates that the ITR's performance is no better than random treatment allocation for the same proportion of patients. Further information explaining the reasons for the PAPE falling within the range of $-0.5$ and $0.5$ can be found in Appendix \ref{app_sec:Generation}. To facilitate a more straightforward comparison of the metric results, we used $P_e = 2 \times \text{PAPE}$ instead of $\text{PAPE}$ to obtain values between $-1$ and $1$ and make it comparable with $C_{\text{fb}}$ and $C_b$. In this paper, we focus exclusively on the situation where the rule $r$ is defined by $\indi{\tau(X)>0}$.

\subsection{Distribution definition}

 In this study, we consider an oracle scenario where the conditional average treatment effects are assumed to be known without estimation error i.e., $\hat{\tau}=\tau$. 
 
 We define $\tau$ on the risk-difference scale. Specifically, for each individual, let $\Pi^0$ and $\Pi^1$
 denote conditional average-specific risks under control and treatment, respectively. Hence, we have $\tau=\Pi^1-\Pi^0$. The distribution of $\Pi^0$ and $\Pi^1$ were specified to be Beta distributions, which are supported on $[0,1]$, allowing the distribution of individualized treatment effects $\tau$ to take values in $[-1,1]$.

Because the same individual-level characteristics may influence both the risk under control and the risk under treatment, the random variables $\Pi^0$ and $\Pi^1$
are expected to be dependent. To define dependent Beta distributions, we relied on the methods introduced in \cite{OLKIN201554}. We first defined a random variable $U=(U_1,U_2,U_3,U_4)$ such that
\begin{align*}
U \sim \operatorname{Dirichlet}(\alpha)
\end{align*}
with $\alpha=\left(\alpha_1, \alpha_2, \alpha_3, \alpha_4\right)$ with $\alpha_i>0 \text { for } i=1, \cdots, 4$ and $\sum_{j=1}^4 U_j=1$.
To define $U$, we first set-up  four independent random variables $G_j$, $j=1, \ldots 4$ such that
\begin{equation}
    G_j  \sim \textnormal{Gamma}(\alpha_j,1)
\end{equation}
and defined 
\begin{equation*}
    U_j=\frac{G_j}{G_1+G_2+G_3+G_4},\qquad j = 1,\dots, 4.
\end{equation*}
Finally, we set $\Pi^1=U_1+U_2$ and $\Pi^0=U_1+U_3$. Given that $\left(U_1+U_2, U_3, U_4\right) \sim \text{Dirichlet} \left(\alpha_1+\alpha_2, \alpha_3, \alpha_4\right)$ and  $\left(U_1+U_3, U_2, U_4\right) \sim \text{Dirichlet} \left(\alpha_1+\alpha_3, \alpha_2, \alpha_4\right)$ \cite{ng2011dirichlet} and that a Dirichlet distribution has beta-distributed marginals, we obtain the following \cite{moschen2023bivariate}:
\begin{align*}
\Pi^0 \sim \operatorname{Beta}\left(\alpha_1+\alpha_3, \alpha_2+\alpha_4\right) \text { and } \Pi^1 \sim \operatorname{Beta}\left(\alpha_1+\alpha_2, \alpha_3+\alpha_4\right)
\end{align*}
together with
\begin{equation*}
    \Cov(\Pi^0,\Pi^1)=\dfrac{\alpha_1 \alpha_4-\alpha_2\alpha_3}{M^2(M+1)}
\end{equation*}
where $M=\sum_{i=1}^4 \alpha_i$.

The potential outcomes are respectively denoted $Y^0$ and $Y^1$. The joint distribution of $(Y^0,Y^1)$ is specified as follows. 
For simplicity, we assumed that conditional on $(\Pi^0,\Pi^1)$ the potential outcomes $(Y^0,Y^1)$ are independent  i.e.
\begin{align*}
    (Y^0,Y^1) \vert (\Pi^0,\Pi^1) \sim \mathrm{Bern}(\Pi^0) \otimes \mathrm{Bern(\Pi^1)}
\end{align*}
where $\mathrm{Bern}(\Pi^t)$ with $t=0,1$ denotes a Bernoulli distribution with parameter $\Pi^t$.

\noindent The parameters for defining distributions were selected to create unimodal or bimodal distributions of $\tau$ with heterogeneous treatment effects, ensuring some individuals benefit from the treatment and others do not. The parameter values were also chosen so that the average treatment effect of the generated distributions was close to $0$ as these scenarios hold the most interest in personalized medicine. Note that
\begin{equation*}
    \Esp\tau=\dfrac{\alpha_2-\alpha_3}{M}.
\end{equation*}
Additionally, mixtures of two Dirichlet distributions with equal weights were used to obtain scenarios with bimodal distributions of $\tau$. The parameters selected for each distribution are listed in Appendix B.
 
\subsection{Metrics calculation}

The values of the metrics for all distributions were computed via numerical integration, using the R statistical software version 4.6 \cite{r_core_team_r_2022}. The specific methods used to compute these metrics are detailed in Appendix \ref{app_sec:Generation}.

\section{Results}\label{sec:res}

\begin{figure}[h]
\centerline{\includegraphics[scale=0.5]{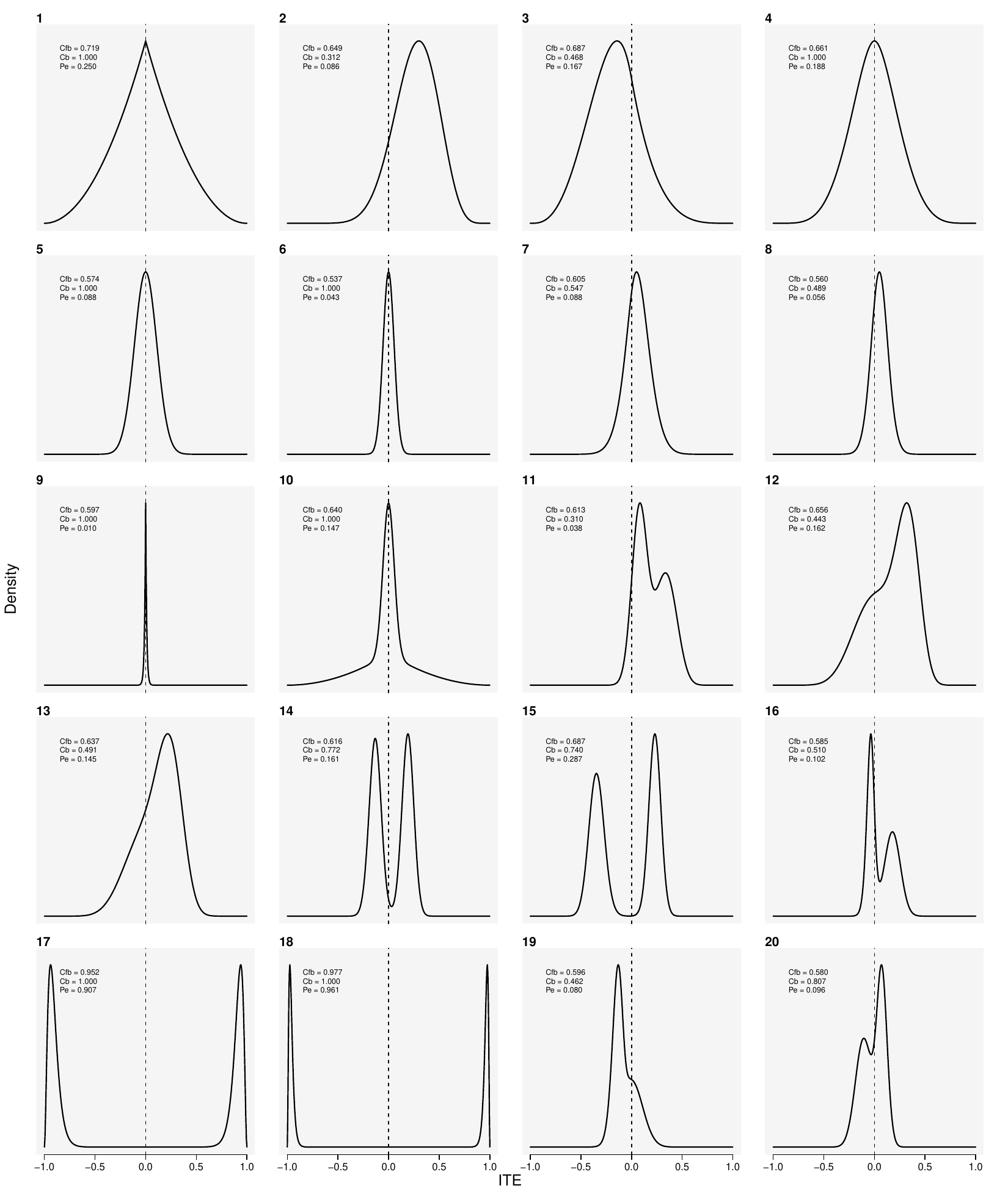}}
\caption{Results of the distributions selected for the analysis.\label{fig:distr}}
\end{figure}

Figure \ref{fig:distr} displays the 20 simulated distributions of individual treatment effects, and Table \ref{tab:metrics} in Appendix B reports the corresponding values of PAPE ($P_e$), the concentration of benefit ($C_b$), and the c-statistic for benefit ($C_{\text{fb}}$). 

Overall, the three metrics varied differently across the distributions indicating that they captured distinct features of the CATE distributions.

Across the 20 distributions, $P_e$ ranged from 0.010 to 0.961, $C_b$ from 0.310 to 1.000, and $C_{\text{fb}}$ from 0.537 to 0.977. The highest values across the three metrics were observed for distributions 17 and 18. These distributions had $C_b = 1.000$, $P_e$ values of 0.907 and 0.961, and $C_{\text{fb}}$ values of 0.952 and 0.977, respectively. Distributions 17 and 18 showed mass concentrated near both extremes of the CATE scale that is close to both large benefit and large non-benefit. These two distributions represent the situations in which discrimination was high regardless of the metric used.

The concentration of benefit reached its maximum value of $1$ in distributions 1, 4, 5, 6, 9, 10, 17, and 18. This pattern is expected because distributions symmetric around zero have an average treatment effect equal to 0; that is, $\mathbb{E}(\tau) = 0$, which leads to $C_b = 1.000$. In contrast, several asymmetric or shifted distributions, including distributions 2, 3, 11, 12, and 13, had lower $C_b$ values, ranging from $0.310$ to $0.491$. Hence, the concentration of benefit can reach its maximal value even when the individualized treatment effect is concentrated around zero

However, the distributions 1, 4, 5, 6, 9 and 10 had very different $P_e$ values. For example, distributions 6 and 9 had $C_b = 1.000$ but $P_e$ values of only 0.043 and 0.010, whereas distribution 1 had $C_b = 1.000$ and $P_e$ = 0.250. Among the visually symmetric distributions, $P_e$ was highest when the distribution placed substantial mass far from zero, especially when both large benefit and large non-benefit were present. This was most evident for distributions 17 and 18. Among the other distributions, distribution 15 had the highest $P_e$, 0.287, followed by distribution 1, 0.250. In contrast, distributions with treatment effects concentrated close to zero, such as distributions 6 and 9, had low $P_e$ despite maximal $C_b$. This suggests that, for balanced distributions, $P_e$ is driven primarily by the magnitude of treatment-effect heterogeneity rather than by the mere presence of individuals on both sides of zero. This result is expected as twice the PAPE associated with the optimal ITR for a mean-zero CATE with a symmetric distribution is equal to the absolute first moment of the CATE.

Bimodality showed a similarly nuanced relationship with $P_e$. With the exception of distributions 17 and 18, the other bimodal distributions exhibited heterogeneous behavior. Distribution 15 had the highest $P_e$ among the remaining distributions, with $P_e = 0.287$, whereas distributions 11 and 20 had much lower $P_e$ values despite being clearly bimodal. This suggests that multimodality alone is insufficient to produce a high $P_e$ value; the location, separation, and relative balance of the modes also play an important role.

The c-statistic for benefit showed a pattern closer to $P_e$ than to $C_b$, although discrepancies remained. Consistent with the patterns observed for $P_e$, distributions 17 and 18 also had the highest $C_{\text{fb}}$ values, while distributions with small treatment-effect magnitudes generally had lower values. However, several distributions with low $P_e$ still had moderate $C_{\text{fb}}$ values. For example, distributions 2, 9, and 11 had $P_e$ values below 0.10 but $C_{\text{fb}}$ values between 0.597 and 0.649. Symmetric distributions with small magnitude of treatment effect such as distributions 5 and 6 have among the lowest values for $C_{\text{fb}}$. This is consistent with the idea that the c-statistic for benefit aims to measure the ability of the model to rank patients according to who benefits more from the treatment.

\begin{figure}[h]
\centerline{\includegraphics[scale=0.5]{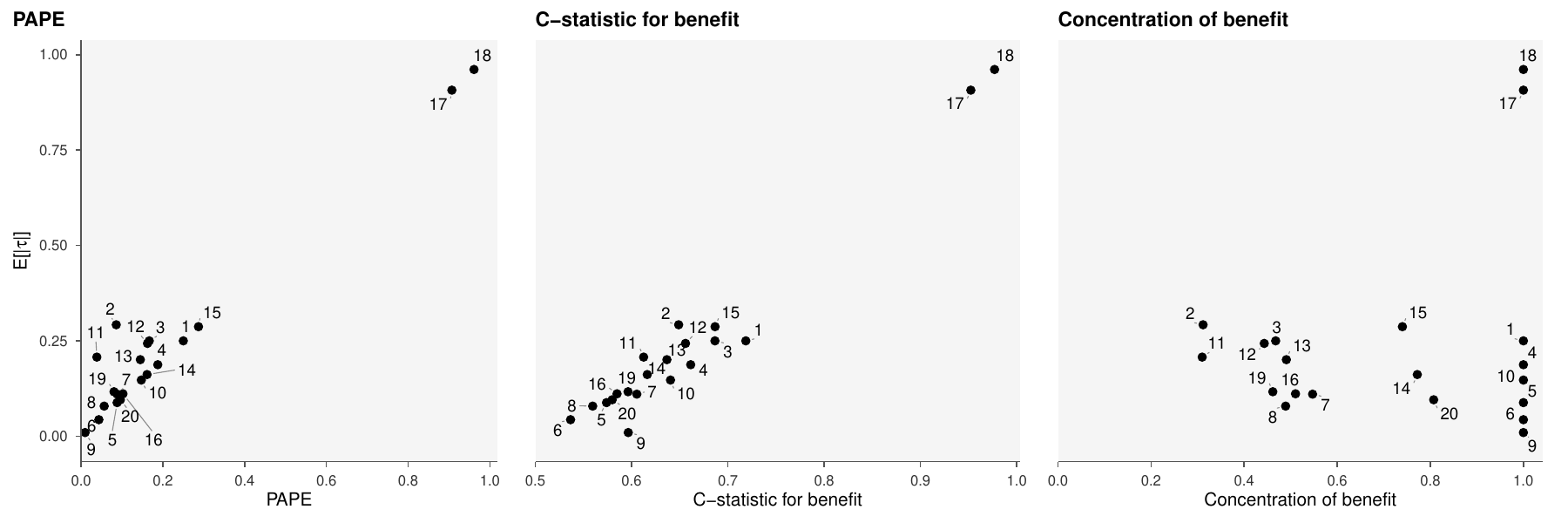}}
\caption{Mean absolute treatment effect versus discrimination metrics across simulated CATE distributions\label{fig:scatter}}
\end{figure}

Figure \ref{fig:scatter} illustrates the influence of the magnitude  of the heterogeneity on the three metrics. The PAPE and the c-statistic for benefit increase with the mean absolute treatment effect while the concentration for benefits exhibits a different behavior which does not seem to reflect meaningful heterogeneity.

Taken together, these findings show that the three metrics do not always agree across distributions and capture complementary aspects of treatment-effect heterogeneity.

\section{Discussion}\label{sec:discuss}
This study examined how different distributions of treatment effects, each with varying levels of heterogeneity, impacted the achievable discrimination. The discrimination of the ITR derived from the oracle CATE was measured with the c-statistic for benefit, concentration of benefit, and population average prescriptive effect. Distributions of treatment effects were defined by employing the difference between two dependent Beta distributions and mixtures of Dirichlet distributions.
\\
Results indicated that favorable performance across all three metrics was observed in distributions that included many individuals with a CATE near $1$ and many with a CATE near $-1$. Alternatively, distributions consisting solely of individuals who experienced either a minor benefit or a minor non-benefit showed poor performance across all metrics. Such distributions are commonly encountered when working with data from randomized controlled trials. Interestingly, the three metrics employed in this study did not demand identical conditions and levels of treatment effect heterogeneity to yield high discrimination results. In general, it was observed that achieving good c-statistic for benefit values was associated with situations in which the individualized treatment effect takes a large range of values. This can be attributed to the role of the c-statistic for benefit, which measures how well we can differentiate individuals with a benefit from those without a benefit. Distributions with wide benefits clearly distinguish these groups, allowing for better discrimination and thus higher $C_{\text{fb}}$ values. Distributions that included equal numbers of individuals benefiting and not benefiting from treatment led to a concentration of benefit of one. Hence, the concentration of benefit can be maximized in cases where $\tau$ has mean zero despite the fact there is no clinically meaningful heterogeneity. In this setting the concentration for benefit is non-informative. The superior performance of an individualized treatment rule i.e., optimal PAPE values, was achieved when distributions encompassed people with a strong benefit and people with a strong non-benefit. The greater heterogeneity present when there are individuals with both strong benefits and strong non-benefits, enhances the effectiveness of the individualized treatment rule, leading to higher PAPE values. The difference between the three metrics suggests that the choice of the metric influences the conclusions drawn from the results. A significantly higher level of heterogeneity is necessary to achieve high c-statistic for benefit values and PAPE compared to what is required for obtaining a high concentration of benefit.
\\
Alternative metrics to evaluate the discrimination of a CATE model include the c-statistic for benefit using 1:1 matching on predicted outcome risk under the control and the model-based c-statistic for benefit \cite{hoogland_evaluating_2022}. However, these were not included in this work, as they assess the same discrimination aspect as the c-statistic for benefit.
\\
Previous simulation studies have looked into the factors contributing to effective discrimination performance. For instance, Rekkas et al. compared diverse methods for constructing ITRs, revealing that larger sample sizes and a moderate average treatment effect ($\text{OR} = 0.8$) were associated with improved c-statistic for benefit values \cite{rekkas_estimating_2023}. Hoogland et al. conducted a simulation study assessing the performance of various discrimination metrics, including the c-statistic for benefit, across different sample sizes \cite{hoogland_evaluating_2022}. They observed that larger sample sizes led to less biased discrimination metrics. 
To our knowledge, no study has specifically examined the discrimination and relative performance of different distributions of treatment effects using the three metrics employed in this study. It might be valuable to undertake a simulation study incorporating diverse sample sizes, distribution shapes, and event rates to discern which parameter exerts the most significant impact on discriminative performance.
\\
One limitation of our calculation of the oracle c-statistic for benefit is that it requires specifying the joint distribution of the two potential outcomes, $(Y^0,Y^1)$. In our data-generating mechanism, we assumed that $Y^0$ and $Y^1$ were conditionally independent given $(\Pi^0,\Pi^1)$. This assumption is convenient for computation but is likely to be unsatisfied in several situations (e.g. a trial comparing different doses of the same drug). The numerical values of $C_{\mathrm{fb}}$ may depend on this unidentifiable joint law.
\\
In conclusion, the presence of heterogeneous treatment effects, whether characterized by variability in the direction or magnitude, did not consistently result in favorable discrimination results. The optimal discrimination depends on the distribution of treatment effects. Furthermore, the choice of metric used to evaluate discrimination also influences the results. In practice, examining the distribution of treatment effects could provide insights into the achievable  discriminative ability. However, the CATEs are unobservable and their predictions are different given the method chosen to develop the model, even in large datasets \cite{bouvier2024machine}.

\section*{Declarations}
\subsection*{Competing interests}
The authors declare that they have no competing interests.
\subsection*{Funding}
FB and RP acknowledge support by the French Agence Nationale de la Recherche as part of the “Investissements d’avenir” program, reference ANR-19-P3IA-0001 (PRAIRIE 3IA Institute).\\
FP acknowledges support by the French Agence Nationale de la Recherche through the project
reference ANR-22-CPJ1-0047-01. 

\subsection*{Authors' contributions}
Study concept and design: FB, FP, and RP. Analysis and interpretation of data: FB, EP, FP, and RP. Drafting of the manuscript: FB and RP. Critical revision of the manuscript for important intellectual content: FB, EP, FP, and RP.
\subsection*{Acknowledgements}
The authors thank François Grolleau and Roch Giorgi for their useful comments and suggestions.

\clearpage
\bibliographystyle{sim}
\bibliography{ITR}

\clearpage
\appendix
\section{Some inequalities for the PAPE} \label{app_sec:PAPE}
Let $r$ be an individualized treatment rule and $p_r = \mathbb{E}(r)$ be the proportion of patients for which treatment is recommended by $r$.
\[
\text{PAPE}=\Esp{(r-p_r) \tau}
\]
$|\text{PAPE}| {\leq} \sqrt{\operatorname{Var}(r)}$ in particular:

$$
\begin{aligned}
|\text{PAPE}| & {\leq} \sqrt{\Esp{(r-p_r)^2}} \sqrt{\Esp{\tau^2}} (\text {Cauchy-Schwarz}) \\
& \leq \sqrt{\Esp{(r-p_r)^2}} \text { since }-1 \leq \tau \leq 1 \\
& \leq \sqrt{\Esp{(r-\Esp{r})^2}} \\
& =\sqrt{\operatorname{Var}(r)} \\
& =\sqrt{p_r\left(1-p_r\right)} \\
&
\end{aligned}
$$
$\sqrt{p_r\left(1-p_r\right)}$ reaches its maximal value at $p_r=\frac{1}{2}$ thus $-\frac{1}{2} \leq \text{PAPE} \leq \frac{1}{2}$\\

There are several straightforward inequalities that help to interpret the PAPE when $r=\indi{\tau(X)>0}$. First, 
\[
\text{PAPE} \geq \min(p_r,1-p_r) \Esp{|\tau|}.
\]
Indeed,
\begin{align*}
    \Esp{(r-p_r) \tau} &= (1-p_r) \Esp{\tau^+ } + p_r \Esp{\tau^- }\\
                       &\geq \min(p_r,1-p_r) \Esp{|\tau|}
\end{align*}
where $\tau^+$ and $\tau^-$ are respectively the positive and negative part of $\tau$. If one further assume  that $\tau$ has a symmetric distribution centered in zero then

\begin{equation*}
    2\text{PAPE}=\Esp{|\tau|}.
\end{equation*}
Moreover, in term of variance
\begin{equation*}
   2\,  \text{PAPE} \geq \operatorname{Var}(\tau).
\end{equation*}
Indeed, in this setting $\text{PAPE}= \Esp{r \tau}$. As $-1 \leq \tau \leq 1$, $\tau r \geq (\tau r)^2$ and since the distribution of $\tau$ is symmetric centered in zero 
\begin{equation*}
    \Esp{(\tau r)^2} = \dfrac{\operatorname{Var}(\tau)}{2}.
\end{equation*}
Hence, $\text{PAPE} \geq \dfrac{\operatorname{Var}(\tau)}{2}$.

\section{Data generating mechanism and computation  of the metrics} \label{app_sec:Generation}
Let
\begin{equation*}
(U_1,U_2,U_3,U_4)\sim \mathrm{Dirichlet}(\alpha_1,\alpha_2,\alpha_3,\alpha_4),
\qquad U_4=1-U_1-U_2-U_3.
\end{equation*}
We define
\begin{equation*}
\Pi^1=U_1+U_2,
\qquad
\Pi^0=U_1+U_3,
\end{equation*}
so that the individual treatment effect is
\begin{equation*}
\tau=\Pi^1-\Pi^0=U_2-U_3.
\end{equation*}

The numerical computations use the following parametrization. Let $G_1, G_2, G_3, G_4$ follow four independent Gamma distributions,
\begin{equation*}
G_j\sim \mathrm{Gamma}(\alpha_j,1),
\qquad j=1,\dots,4.
\end{equation*}
and define
\begin{equation*}
    U_j=\frac{G_j}{G_1+G_2+G_3+G_4},\qquad j = 1,\dots, 4.
\end{equation*}
Then $U\sim \mathrm{Dirichlet}(\alpha)$. Define
\begin{equation*}
S=U_2+U_3,
\qquad
V=\frac{U_2}{U_2+U_3},
\qquad
T=\frac{U_1}{U_1+U_4}.
\end{equation*}
Equivalently,
\begin{equation*}
U_1=(1-S)T, \qquad
U_2=SV, \qquad
U_3=S(1-V), \qquad
U_4=(1-S)(1-T).
\end{equation*}
By the beta-gamma algebra,
\begin{equation*}
S\sim \mathrm{Beta}(\alpha_2+\alpha_3,\alpha_1+\alpha_4),\qquad
V\sim \mathrm{Beta}(\alpha_2,\alpha_3),\qquad
T\sim \mathrm{Beta}(\alpha_1,\alpha_4),
\end{equation*}
and $S,V,T$ are mutually independent. Therefore,
\begin{equation*}
\Pi^1 = (1-S)T+SV,\qquad
\Pi^0 = (1-S)T+S(1-V), \qquad
\tau=\Pi^1-\Pi^0=S(2V-1).
\end{equation*}

\paragraph{PAPE.}

The PAPE can be written as
\begin{align*}
\mathrm{PAPE}
&= \Esp{\left\{\indi{\tau>0} - \Prob{\tau>0}\right\}\,\tau}\\
&= \Esp{\tau\indi{\tau>0}} - \Prob{\tau>0}\Esp{\tau}.
\end{align*}

Since $\tau=S(2V-1)$ and $S>0$, we have
\begin{equation*}
\tau>0
\quad\Longleftrightarrow\quad
V>\frac12.
\end{equation*}
Thus,
\begin{equation*}
\Prob{\tau>0} = \Prob{V>\frac12},
\qquad
V\sim \mathrm{Beta}(\alpha_2,\alpha_3).
\end{equation*}
Moreover,
\begin{equation*}
\Esp{\tau}
=
\Esp{S}\Esp{2V-1}
=
\frac{\alpha_2-\alpha_3}
{\alpha_1+\alpha_2+\alpha_3+\alpha_4}.
\end{equation*}
Finally,
\begin{align*}
\Esp{\tau\indi{\tau>0}}
&= \Esp{S(2V-1)\indi{V>1/2}}\\
&= \Esp{S}\Esp{(2V-1)\indi{V>1/2}},
\end{align*}
because $S$ and $V$ are independent. Hence all terms needed for PAPE are obtained from beta distribution functions.

\paragraph{Concentration of benefit.}

Recall that the concentration of benefit is defined as
\begin{equation*}
C_b=
\begin{cases}
    1 - \frac{\Esp{\tau}}{\Esp{\max(\tau,\tau')}},\text{ if }\Esp{\tau} \geq 0,\\
    1 - \frac{\Esp{\tau}}{\Esp{\min(\tau,\tau')}}, \text{ otherwise.}
\end{cases}
\end{equation*}
where $\tau'$ is an independent copy of $\tau$.

For simplicity we only compute $\Esp{\max(\tau, \tau')}$ since
\begin{equation*}
    \min(\tau,\tau') + \max(\tau,\tau') =\tau+\tau' \implies \Esp{\min(\tau,\tau')} = 2\Esp{\tau} - \Esp{\max(\tau,\tau')}
\end{equation*}

Let $F_\tau$ denote the cumulative distribution function of $\tau$. Since $\tau\in[-1,1]$, we use
\begin{equation*}
\Esp{\max(\tau,\tau')} = 1 - \int_{-1}^1 F_\tau(t)^2\, dt.
\end{equation*}
It remains to compute $F_\tau$. Using
\begin{equation*}
    \tau=S(2V-1),
\end{equation*}
we compute the CDF by one-dimensional numerical integration.

For \(t<0\),
\begin{align*}
    F_\tau(t) &= \Prob{S(2V-1) \leq t}\\
    &= \Prob{S \geq \frac{t}{2V-1}},\text{ $2V-1$ has the same sign as $t<0$.}\\
    &= \Esp{\Prob{S \geq \frac{t}{2V-1} \mid V}}\\
    &= \int \left[ 1-F_S\left(\frac{t}{2v-1}\right) \right] dF_V(v)\\
    &= \int_0^{(1+t)/2} \left[ 1-F_S\left(\frac{t}{2v-1}\right) \right] dF_V(v),\text{ $S\leq 1$ and $V\geq0$.}
\end{align*}
where
\begin{equation*}
    S\sim \mathrm{Beta}(\alpha_2+\alpha_3,\alpha_1+\alpha_4),
    \qquad
    V\sim \mathrm{Beta}(\alpha_2,\alpha_3).
\end{equation*}

For $0\leq t<1$,
\begin{align*}
    F_\tau(t) &= \Prob{S(2V-1) \leq t}\\
    &= \Esp{\Prob{S \leq \frac{t}{2V-1} \mid V}},\text{ $2V-1$ has the same sign as $t\geq0$.}\\
    &= \int F_S\left(\frac{t}{2v-1}\right) dF_V\\
    &= F_V\left(\frac{t+1}{2}\right) + \int_{(t+1)/2}^1 F_S\left(\frac{t}{2v-1}\right) dF_V,\text{ $S\leq 1$ and $V\leq1$.}
\end{align*}
Once $F_\tau$ is evaluated on a grid over $[-1,1]$, we compute
\begin{equation*}
C_b =
\begin{cases}
    1- \frac{\Esp{\tau}}{1-\int_{-1}^{1}F_\tau(t)^2\,dt}, \text{ if }\Esp{\tau}\geq 0 \\
    1- \frac{\Esp{\tau}}{2\Esp{\tau} -1 +\int_{-1}^{1}F_\tau(t)^2\,dt}, \text{ otherwise.}
\end{cases}
\end{equation*}

\paragraph{C-statistic for benefit.}

The c-statistic for benefit is defined as
\begin{equation*}
C_{\mathrm{fb}} = \Prob{\tau>\tau'\mid B>B'} = \frac{\Prob{B>B',\tau>\tau'}}{\Prob{B>B'}},
\end{equation*}
where $B=Y_1-Y_0$, and where primed quantities denote an independent copy of the same random variables. We need to compute the joint probability $\Prob{B>B',\tau>\tau'}$ and the marginal probability $\Prob{B>B'}$.

Conditional on $\Pi^0,\Pi^1$, we assume
\begin{equation*}
Y_1\sim \mathrm{Bernoulli}(\Pi^1),
\qquad
Y_0\sim \mathrm{Bernoulli}(\Pi^0),
\end{equation*}
with $Y_1$ and $Y_0$ conditionally independent. Therefore,
\begin{align*}
&\Prob{B=-1\mid \Pi^0,\Pi^1} = \Pi^0(1-\Pi^1),\\
&\Prob{B=0\mid \Pi^0,\Pi^1} = \Pi^0\Pi^1+(1-\Pi^0)(1-\Pi^1),\\
&\Prob{B=1\mid \Pi^0,\Pi^1} = (1-\Pi^0)\Pi^1.
\end{align*}

For two independent individuals, define
\begin{equation*}
f(\Pi^0,\Pi^{0'},\Pi^1,\Pi^{1'}) = \Prob{B>B'\mid \Pi^0,\Pi^{0'},\Pi^1,\Pi^{1'}}.
\end{equation*}
Then
\begin{equation*}
f(\Pi^0,\Pi^{0'},\Pi^1,\Pi^{1'}) =[1-\Pi^0(1-\Pi^1)]\,\Pi^{0'}(1-\Pi^{1'})
\; + \;
(1-\Pi^0)\Pi^1  \,\left[\Pi^{0'}\Pi^{1'}+(1-\Pi^{0'})(1-\Pi^{1'})\right].
\end{equation*}
The first term corresponds to $B\in\{0,1\}$ and $B'=-1$. The second term corresponds to $B=1$ and $B'=0$.

Using the tower rule,
\begin{align*}
\Prob{B>B',\tau>\tau'}
&= \Esp{\Esp{\indi{B>B',\tau>\tau'}\mid \Pi^0,\Pi^{0'},\Pi^1,\Pi^{1'}}}\\
&= \Esp{f(\Pi^0,\Pi^{0'},\Pi^1,\Pi^{1'})\indi{\tau>\tau'}}.
\end{align*}
Therefore,
\begin{equation*}
C_{\mathrm{fb}} = \frac{ \Esp{f(\Pi^0,\Pi^{0'},\Pi^1,\Pi^{1'})\indi{\tau>\tau'}} }{ \Prob{B>B'} }.
\end{equation*}

The denominator can be computed from the marginal distribution of $B$. Let
\begin{equation*}
p_-=\Prob{B=-1}, \qquad p_0=\Prob{B=0}, \qquad p_+=\Prob{B=1}.
\end{equation*}
Marginally,
\begin{align*}
p_- &= \Esp{\Pi^0(1-\Pi^1)},\\
p_0 &= \Esp{\Pi^0\Pi^1+(1-\Pi^0)(1-\Pi^1)},\\
p_+ &= \Esp{(1-\Pi^0)\Pi^1}.
\end{align*}
Since $B$ and $B'$ are iid,
\begin{equation*}
\Prob{B>B'} = p_+(p_0+p_-)+p_0p_-.
\end{equation*}

It remains to compute the numerator. This integral is simplified using the variables $S,V,T$. Since
\begin{equation*}
\tau=S(2V-1),
\end{equation*}
the indicator does not depend on $T$ or $T'$. Conditional on $S=s$ and $V=v$, define
\begin{align*}
m_0(s,v)&=\Esp{\Pi^0\mid S=s,V=v},\\
m_1(s,v)&=\Esp{\Pi^1\mid S=s,V=v},\\
m_{01}(s,v)&=\Esp{\Pi^0\Pi^1\mid S=s,V=v}.
\end{align*}
Using
\begin{equation*}
\Esp{T} = \frac{\alpha_1}{\alpha_1+\alpha_4}, \qquad
\Esp{T^2} = \frac{\alpha_1(\alpha_1+1)}{(\alpha_1+\alpha_4)(\alpha_1+\alpha_4+1)},
\end{equation*}
we obtain
\begin{align*}
m_0(s,v) &= (1-s)\Esp{T}+s(1-v),\\
m_1(s,v) &= (1-s)\Esp{T}+sv,\\
m_{01}(s,v) &= (1-s)^2\Esp{T^2} + s(1-s)\Esp{T} + s^2v(1-v).
\end{align*}

Now define
\begin{align*}
A_1(s,v) &= 1-m_0(s,v)+m_{01}(s,v),\\
A_2(s,v) &= m_1(s,v)-m_{01}(s,v),\\
B_1(s,v) &= m_0(s,v)-m_{01}(s,v),\\
B_2(s,v) &= 1-m_0(s,v)-m_1(s,v)+2m_{01}(s,v).
\end{align*}
Then, conditional on $(S,V,S',V')=(s,v,s',v')$,
\begin{equation*}
\Esp{f(\Pi^0,\Pi^{0'},\Pi^1,\Pi^{1'}) \mid S,V,S',V'} = h(s,v,s',v'),
\end{equation*}
where
\begin{equation*}
h(s,v,s',v') = A_1(s,v)B_1(s',v') + A_2(s,v)B_2(s',v').
\end{equation*}
Hence
\begin{equation*}
\Prob{B>B',\tau>\tau'} = \int h(s,v,s',v') \indi{s(2v-1)>s'(2v'-1)} dF_S(s)dF_V(v)dF_S(s')dF_V(v').
\end{equation*}
Finally,
\begin{equation*}
C_{\mathrm{fb}} = \frac{
\int h(s,v,s',v') \indi{s(2v-1)>s'(2v'-1)} dF_S(s)dF_V(v)dF_S(s')dF_V(v')
}{
\Prob{B>B'}
}.
\end{equation*}
The numerator is the four-dimensional integral evaluated numerically, while the denominator is computed from the marginal probabilities $p_-,p_0,p_+$.

\paragraph{Extension to a two-component Dirichlet mixture.}
Now suppose
\begin{equation*}
U
\sim
w\,\mathrm{Dirichlet}(\alpha^{(1)})
+
(1-w)\,\mathrm{Dirichlet}(\alpha^{(2)}).
\end{equation*}
Let
\begin{equation*}
w_1=w,
\qquad
w_2=1-w.
\end{equation*}
For each component $k\in\{1,2\}$, the previous formulas apply with
\begin{equation*}
\alpha=\alpha^{(k)}.
\end{equation*}

For PAPE, the required quantities are averaged over the mixture:
\begin{align*}
\Esp{\tau} &= \sum_{k=1}^2 w_k\Esp[k]{\tau },\\
\Prob{\tau>0} &= \sum_{k=1}^2 w_k\Prob[k]{\tau>0},\\
\Esp{\tau\indi{\tau>0}} &= \sum_{k=1}^2 w_k \Esp[k]{\tau\indi{\tau>0}}.
\end{align*}
Then
\begin{equation*}
\mathrm{PAPE} = \Esp{\tau\indi{\tau>0}} - \Prob{\tau>0}\Esp{\tau}.
\end{equation*}

For CB, the mixture CDF is
\begin{equation*}
F_\tau(t)
=
\sum_{k=1}^2 w_kF_{\tau,k}(t).
\end{equation*}
Therefore,
\begin{equation*}
\Esp{\max(\tau_1,\tau_2)} = 1 - \int_{-1}^1 F_\tau(t)^2\,dt
\end{equation*}
and
\begin{equation*}
\mathrm{CB}
=
1-
\frac{\Esp{\tau}}
{\Esp{\max(\tau_1,\tau_2)}}.
\end{equation*}

For the c-statistic for benefit, the two individuals are drawn independently from the mixture. Thus the component pair $(k,\ell)$ has weight $w_kw_\ell$. The numerator is
\begin{equation*}
\Prob{B>B',\tau>\tau'} =
\sum_{k=1}^2 \sum_{\ell=1}^2 w_k w_\ell
\Esp[k,\ell]{f(\Pi^0,\Pi^{0'},\Pi^1,\Pi^{1'}) \indi{\tau>\tau'}}.
\end{equation*}
The denominator is computed from the marginal distribution of \(B\) under the mixture. For \(b\in\{-,0,+\}\),
\begin{equation*}
p_b=\sum_{k=1}^2 w_k p_{b,k}.
\end{equation*}
Since \(B\) and \(B'\) are iid,
\begin{equation*}
\Prob{B>B'}
=
p_+(p_0+p_-)+p_0p_-.
\end{equation*}
Hence
\begin{equation*}
C_{\mathrm{fb}} =
\frac{
    \sum_{k=1}^2 \sum_{\ell=1}^2 w_k w_\ell \Esp[k,\ell]{f(\Pi^0, \Pi^{0'},\Pi^1,\Pi^{1'}) \indi{\tau>\tau'}}
}{
    p_+(p_0+p_-)+p_0p_-
}.
\end{equation*}
Equivalently, the numerator is a weighted sum of four deterministic four-dimensional integrals, one for each pair of Dirichlet components, and the ratio is formed only after summing these terms.

\section*{Appendix B}\label{app:table}

\begin{sidewaystable}[!h]

\caption{\label{tab:metrics}Metrics table}
\centering
\fontsize{8}{10}\selectfont
\begin{tabular}[t]{rrrrrrrrrrrrrrrrrrrr}
\toprule
\multicolumn{1}{c}{ } & \multicolumn{4}{c}{$a$} & \multicolumn{4}{c}{$b$} & \multicolumn{1}{c}{ } & \multicolumn{10}{c}{Metrics} \\
\cmidrule(l{3pt}r{3pt}){2-5} \cmidrule(l{3pt}r{3pt}){6-9} \cmidrule(l{3pt}r{3pt}){11-20}
id & $a_1$ & $a_2$ & $a_3$ & $a_4$ & $b_1$ & $b_2$ & $b_3$ & $b_4$ & mixture & PAPE & CB & C-stat benefit & $\Esp{\tau}$ & $\mathrm{Var}(\tau)$ & $\Prob{\tau > 0}$ & $\Esp{\tau \,\indi{\tau>0}}$ & CB denominator & C-stat numerator & C-stat denominator\\
\midrule
1 & 1 & 1 & 1 & 1 & -- & -- & -- & -- & no & 0.250 & 1.000 & 0.719 & 0.000 & 0.100 & 0.500 & 0.125 & 0.179 & 0.225 & 0.312\\
2 & 2 & 8 & 4 & 1 & -- & -- & -- & -- & no & 0.086 & 0.312 & 0.649 & 0.267 & 0.046 & 0.887 & 0.279 & 0.387 & 0.198 & 0.305\\
3 & 2 & 1 & 2 & 1 & -- & -- & -- & -- & no & 0.167 & 0.468 & 0.687 & -0.167 & 0.067 & 0.250 & 0.042 & -0.313 & 0.210 & 0.306\\
4 & 2 & 2 & 2 & 2 & -- & -- & -- & -- & no & 0.188 & 1.000 & 0.661 & 0.000 & 0.056 & 0.500 & 0.094 & 0.133 & 0.207 & 0.312\\
5 & 10 & 10 & 10 & 10 & -- & -- & -- & -- & no & 0.088 & 1.000 & 0.574 & 0.000 & 0.012 & 0.500 & 0.044 & 0.062 & 0.179 & 0.312\\
\addlinespace
6 & 42 & 42 & 42 & 42 & -- & -- & -- & -- & no & 0.043 & 1.000 & 0.537 & 0.000 & 0.003 & 0.500 & 0.022 & 0.031 & 0.168 & 0.312\\
7 & 2 & 3 & 2 & 10 & -- & -- & -- & -- & no & 0.088 & 0.547 & 0.605 & 0.059 & 0.016 & 0.688 & 0.085 & 0.130 & 0.165 & 0.273\\
8 & 9 & 7 & 5 & 19 & -- & -- & -- & -- & no & 0.056 & 0.489 & 0.560 & 0.050 & 0.007 & 0.726 & 0.064 & 0.098 & 0.169 & 0.302\\
9 & 100 & 1 & 1 & 1 & -- & -- & -- & -- & no & 0.010 & 1.000 & 0.597 & 0.000 & 0.000 & 0.500 & 0.005 & 0.007 & 0.022 & 0.037\\
10 & 1 & 1 & 1 & 1 & 40 & 40 & 40 & 40 & yes & 0.147 & 1.000 & 0.640 & 0.000 & 0.052 & 0.500 & 0.074 & 0.116 & 0.200 & 0.312\\
\addlinespace
11 & 30 & 28 & 21 & 13 & 12 & 24 & 9 & 1 & yes & 0.038 & 0.310 & 0.613 & 0.201 & 0.025 & 0.920 & 0.204 & 0.291 & 0.186 & 0.304\\
12 & 1 & 10 & 10 & 1 & 12 & 24 & 9 & 1 & yes & 0.162 & 0.443 & 0.656 & 0.163 & 0.053 & 0.748 & 0.203 & 0.293 & 0.204 & 0.311\\
13 & 2 & 20 & 10 & 10 & 1 & 10 & 10 & 1 & yes & 0.145 & 0.491 & 0.637 & 0.119 & 0.042 & 0.735 & 0.160 & 0.234 & 0.199 & 0.313\\
14 & 28 & 39 & 16 & 38 & 35 & 8 & 19 & 21 & yes & 0.161 & 0.772 & 0.616 & 0.029 & 0.030 & 0.507 & 0.095 & 0.126 & 0.193 & 0.313\\
15 & 17 & 4 & 25 & 15 & 26 & 32 & 9 & 33 & yes & 0.287 & 0.740 & 0.687 & -0.057 & 0.087 & 0.500 & 0.115 & -0.220 & 0.220 & 0.321\\
\addlinespace
16 & 30 & 3 & 6 & 36 & 24 & 25 & 12 & 13 & yes & 0.102 & 0.510 & 0.585 & 0.068 & 0.016 & 0.565 & 0.090 & 0.139 & 0.182 & 0.311\\
17 & 1 & 1 & 40 & 1 & 1 & 40 & 1 & 1 & yes & 0.907 & 1.000 & 0.952 & 0.000 & 0.826 & 0.500 & 0.453 & 0.468 & 0.274 & 0.288\\
18 & 1 & 1 & 100 & 1 & 1 & 100 & 1 & 1 & yes & 0.961 & 1.000 & 0.977 & 0.000 & 0.924 & 0.500 & 0.481 & 0.487 & 0.262 & 0.268\\
19 & 50 & 1 & 10 & 1 & 10 & 10 & 10 & 10 & yes & 0.080 & 0.462 & 0.596 & -0.073 & 0.013 & 0.250 & 0.022 & -0.135 & 0.153 & 0.256\\
20 & 36 & 16 & 9 & 33 & 31 & 16 & 24 & 5 & yes & 0.096 & 0.807 & 0.580 & -0.015 & 0.013 & 0.512 & 0.040 & -0.080 & 0.177 & 0.305\\
\bottomrule
\end{tabular}
\end{sidewaystable}

\end{document}